 \documentclass[11pt]{article}

 \pdfoutput=1

\usepackage[english]{babel}

\usepackage[round]{natbib} 

\usepackage{graphicx,siunitx,xspace}
\usepackage{amsmath}
\usepackage{amssymb}
\usepackage{hyperref}
\usepackage{enumitem}
\usepackage{color}
\usepackage{enumitem}
\usepackage{float}
\usepackage{amsfonts,bm}
\usepackage{epstopdf}
\usepackage[mathscr]{euscript}

  \usepackage[normalem]{ulem}
  \usepackage{textcomp}

\usepackage[utf8]{inputenc}
\usepackage[T1]{fontenc}
\usepackage{pifont} 

\usepackage{chemformula}

\usepackage{float}
\usepackage{ifthen}
\begin{document}


\begin{center}
{\bf \Large An English translation of the paper of Max 
\citet[][]{Planck_history_quantum_1943} about}
\\ \vspace*{2mm}
{\bf \Large ``\,\underline{Zur Geschichte der Auffindung des physikalischen Wirkungsquantums}\,''}
\\ \vspace*{2mm}
{\bf \Large or: ``\,\underline{The history of the discovery of the physical quantum of action}\,''}
\\ \vspace*{2mm}
{\bf \Large to provide a readable version of the German content.}
\\ \vspace*{2mm}
{\bf\color{red} \large Translated by Dr.Hab. Pascal Marquet 
}
\\ \vspace*{2mm}
{\bf\bf\color{red}  \large Possible contact at: 
    pascalmarquet@yahoo.com}
    \\
{\bf\bf\color{red} 
    Web Google-sites:
    \url{https://sites.google.com/view/pascal-marquet}
    \\ ArXiv: 
    \url{https://arxiv.org/find/all/1/all:+AND+pascal+marquet/0/1/0/all/0/1}
    \\ Research-Gate:
    \url{https://www.researchgate.net/profile/Pascal-Marquet/research}
}
\\ \vspace*{-1mm}
\end{center}

\hspace*{65mm} Version-1 / \today
\vspace*{-6mm}

\bibliographystyle{ametsoc2014}
\bibliography{Book_FAQ_Thetas_arXiv}
\vspace*{0mm}

Uncertainties/alternatives in the translation are indicated {\color{red} (in red)} with {\it\color{red} italic terms}, together with some additional footnotes (indicated with {\it\color{red} P. Marquet)}. 
Moreover, I have added some highlight (shown as \dashuline{\,dashing text}), in particular about the \dashuline{integration constant issue for the entropy}.

Do not hesitate to contact me in case of mistakes or any trouble in the English translation from the German text.

\vspace*{-2mm} 
\begin{center}
--------------------------------------------------- 
\end{center}
\vspace*{-11mm}

  \tableofcontents

\vspace*{-1mm} 
\begin{center}
----------------------------------------------------------------------------------------------------------------------- 
\end{center}
\vspace*{-5mm}


\begin{center}
{\bf \Large ``\,\underline{The history of the discovery of the physical quantum of action}\,''}
\end{center}
\vspace*{-6mm}

\begin{center}
{\bf \Large by Max Planck, Berlin (1943).}
\end{center}
\vspace*{-3mm}

Since the appearance of the elementary quantum of action marks the beginning of a new epoch in physical science, I feel the need and the obligation towards the physicists of a later generation to describe in a summarising presentation, to the best of my knowledge, the repeatedly tortuous path by which I arrived at the calculation of this universal constant, as it is reflected in my memory.
\vspace*{-3mm}

\section{\underline{Section-I} {\color{red}{\it(Thermodynamics and entropy)}} (p.153-154)} 
\vspace*{-2mm}

For this purpose, I must first go back a little further, to my university years.
What has always interested me most in physics were the great general laws that have significance for all natural processes, regardless of the properties of the bodies involved in the processes and of the ideas one forms about their structure.
I was therefore particularly fascinated by the two laws of thermodynamics. 
However, while the first law, the law of conservation of energy, has a very simple and easily comprehensible meaning and therefore requires no special explanation, the correct understanding of the second law requires detailed study.
I learnt this principle {\color{red}{\it(second law)}} in my last year of study (1878) by reading the writings of R. Clausius, which particularly attracted me anyway because of the excellent clarity and persuasiveness of the 
language$\,$\footnote{$\:${R. Clausius, Die mechanische Wärmetheorie. 2. Aufl. {\bf I.} {\color{red}{\it[\:The mechanical theory of heat. 2nd edition\,]}} Braunschweig: Friedrich Vieweg und Sohn 1876}.}.

Clausius derived the proof of his second law from the hypothesis that ``\,\dashuline{\,heat does not pass automatically from a colder to a warmer body\,}.\,''
However, this hypothesis requires a special explanation, because it is not only intended to express the fact that heat does not pass directly from a colder to a warmer body, but also that it is not possible in any way to transfer heat from a colder body to a warmer body, for example through a suitably devised circular-{\color{red}{\it(cyclic)}} process, without any other change occurring in nature that serves as compensation, which has the property that it cannot be reversed without leaving behind another permanent change.
Only if one makes this \dashuline{\,further assertion of the hypothesis a prerequisite\,} is it possible \dashuline{\;to provide the general proof of the second law\,}. The multiple attacks that Clausius' proof has received are largely based on a misunderstanding of the complete content of his hypothesis.

In an effort to gain as much clarity as possible on this point, I came up with a formulation of the hypothesis that seemed to me to be simpler and more convenient.
It reads: ``\,\dashuline{\,The process of heat conduction cannot be completely reversed in any way\,}.\,''
This expresses the same thing as the Clausius version without the need for any special explanation. You just have to pay careful attention to the words ``\,in any way\,'' and ``completely.''
They want to say that in the attempt to reverse the process, one may use any means at all, mechanical, thermal, electrical, chemical, only on the condition that after the end of the process used, the means used are again in exactly the same state as at the beginning, when they were used. 
I called a process that cannot be completely reversed in any way ``\,\dashuline{natural}\,,'' today it is called ``\,\dashuline{\,irreversible}.''
But the mistake that one commits by interpreting Clausius' theorem too narrowly, and which I have tried to combat all my life, has, it seems, still not been eradicated for the time being. Because to this day I still encounter the following definition of irreversibility in addition to the above one: ``\,Irreversible is a process that cannot run in the opposite direction.'' This is not sufficient. Because from the outset, it is quite conceivable that a process that cannot run in the opposite direction can be completely reversed in some way by suitably constructed machinery. It is precisely this deeper meaning of irreversibility that makes the second law significant not only for thermal phenomena, but for all natural processes.

According to the above definition, all processes in nature fall into two classes: reversible and irreversible processes (I used to say neutral and natural processes), depending on whether or not they can be completely reversed in some way.
From this it follows, which is now the essential point, that the decision as to whether a natural process is irreversible or reversible depends only on the nature of the initial state and the final state.
There is no need to know anything about the nature and course of the process. 
In the first case, that of irreversible processes, the final state is in a certain sense distinguished from the initial state, and  nature has a greater ``\,preference\,'' for it, so to speak. In the second case, that of reversible processes, the two states have equal status. Clausius' entropy was found to be a measure of the magnitude of this preference, and the meaning of the second law is the law that in every natural process the sum of the entropies of all bodies involved in the process increases, and in the limiting case, for a reversible process, remains unchanged. I used the above statements for my Munich doctoral 
dissertation$\,$\footnote{$\:${\color{red}{\it(M. Planck)}} {Über den zweiten Hauptsatz der mechanischen Wärmetheorie. München {\color{red}{\it[\,On the second law of mechanical heat theory. Munich\,]}} Th. Ackermann, 1879. / {\color{red}\it See the English translation I have made of this Thesis report of 1879, and uploaded in arXiv and Zenodo (P. Marquet).}}}.
The impression this {\color{red}{\it(Doctoral dissertation)}} paper had on the physical public at the time was zero. 
As I know from conversations with \dashuline{\,my university teachers\,}, none of them had any understanding of its content. They probably only let it pass as a dissertation because they knew me from my other work in the practical course in physics and in the mathematics seminar. But even among physicists who were closer to the subject, I found no interest, let alone applause. 
\dashuline{\,Helmholtz\,} had probably not read the paper at all, 
and \dashuline{\,Kirchhoff\,} explicitly rejected its content, stating that the concept of entropy, whose magnitude can only be measured and therefore defined by a reversible process, should not be applied to irreversible processes. 
I was unable to get close to \dashuline{\,Clausius}, who was very reserved in his personal relationship. An attempt I made once to introduce myself to him in Bonn was unsuccessful, because I didn't meet him at home.
However, such experiences did not prevent me, deeply imbued with the importance of this task, from continuing the study of entropy, which I regarded, along with energy, to be the most important property of a physical entity. Since its maximum denotes the final equilibrium, knowledge of entropy gave rise to all the laws of physical and chemical equilibrium. In the following years I carried this out in detail in various works, first for changes in the state of matter, then for gas mixtures and finally for solutions. Fruitful results were obtained everywhere. Unfortunately, however, as I only discovered later, the great American theorist John Willard Gibbs, who had formulated the same theorems earlier, even in a more general version, had beaten me to 
it$\,$\footnote{$\:${J.W. Gibbs, Transactions of the Connecticut Academy 1873, 1876, 1878. Deutsche Übersetzung von Wilh. Ostwald, mit dem Titel Thermodynamische Studien. Leipzig: W. Engelmann 1892 {\color{red}{\it[\,J.W. Gibbs, Transactions of the Connecticut Academy 1873, 1876, 1878. German translation by Wilh. Ostwald, entitled ``\,Thermodynamic Studies.''\,]}}}}, so that I was not particularly successful in this field either.
\vspace*{-3mm}


\section{\underline{Section-II} {\color{red}{\it(Heuristic formulas for Radiation)}} (p.154-157)} 
\vspace*{-2mm}

On the other hand, I discovered new territory in the area of radiant heat. As early as 1860 G. Kirchhoff had taught the theorem that in an evacuated cavity bounded by totally reflecting walls and containing any number of emitting and absorbing bodies, a stationary state of radiation develops over time through irreversible processes, which depends on a single variable, the temperature $T$ common to all bodies. It is the same state of radiation that prevails in a vacuum when the surrounding walls are black and have the temperature in question. It corresponds to a very specific distribution of radiant energy over the individual oscillation numbers $\nu$ of the spectrum.
This so-called normal energy distribution is therefore represented by a universal function of $T$ and $\nu$ that does not depend on any material, and since I am convinced that the more comprehensive a law of nature is, the simpler it is, the more tempting it seemed to me to search for this function.
The direct way to do this was to use Maxwell's electromagnetic theory of light, which had achieved 
final victory a few years earlier thanks to Hertz's great discovery. So I imagined the evacuated cavity filled with 
electrically oscillating bodies that radiate and absorb energy and, 
since their nature is not important, I chose bodies of the simplest possible nature, namely linear resonators or oscillators of a certain natural frequency $\nu$ and weak damping caused only by radiation. My hope was that for an arbitrarily assumed initial state of this structure, the application of Maxwell's theory would lead to irreversible radiation processes, which would have to lead to a stationary state, that of thermodynamic equilibrium, in which the cavity radiation has the desired normal energy distribution corresponding to the radiation of the black body.

Accordingly, I first began an investigation into the absorption and emission of electric waves through resonance$\,$\footnote{$\:${\color{red}{\it(M. Planck)}} Sitzgsber. Berl. Akad. Wiss. 21. III. 1895 {\color{red}[\,Sitzungsberichte der Deutschen/K\"oniglich Preussischen} {\color{red}Akademie der Wissenschaften zu Berlin / {\it Reports of the meetings of the German/Prussian Academy of Sciences in Berlin\,]}}}. 
I was of the opinion that the interaction between an energy-absorbing and emitting oscillator excited by an electrodynamic wave and the wave exciting it represents an irreversible process$\,$\footnote{$\:${{\color{red}\it(M. Planck)}} Sitzgsber, Berl. Akad. Wiss. vom 4. II. 1897, S. 59.}. 
However, this opinion, expressed in such general terms, is erroneous, as L. Boltzmann quickly pointed out$\,$\footnote{$\:$L. Boltzmann, Sitzgsber. Berl. Akad. Wiss. vom 17. VI, 1897.}.
This is because the whole process can just as easily take place in the opposite direction. You only need to reverse the sign of all magnetic field strengths at some point, while retaining the electric field strengths. The oscillator then absorbs the energy emitted in concentric spherical waves in the same spherical waves and emits the energy absorbed by the exciting radiation. There can therefore be no question of irreversibility in such a process.

In order to make any progress at all in the theory of thermal radiation, it is therefore necessary to introduce a limiting condition which excludes from the outset such singular processes which probably never occur in nature, such as concentric inwardly directed spherical waves, and thus also the possibility of a simultaneous reversal of the sign of all magnetic field strengths. I took this step by establishing the hypothesis of ``\,natural radiation\,''$\,$\footnote{$\:${\color{red}{\it(M. Planck)}} Sitzgsber. Berl. Akad. Wiss. vom 7. VII, 1898.}, the content of which boils down to the fact that the individual harmonic partial oscillations that make up a thermal radiation wave are completely incoherent.
On the basis of this hypothesis, I then developed the laws of radiation processes in an evacuated cavity filled with linear oscillators with certain natural frequencies and weak damping, first for a hollow sphere with such an oscillator at its centre, because the differential equations of the process can then be easily integrated, and then in summary for the general case of any cavity with any number of oscillators$\,$\footnote{$\:${\color{red}{\it(M. Planck)}} Sitzgsber. Berl. Akad. Wiss. vom 18. V. 1899.}. 
As a result of this investigation, the theorem emerged that the interaction of an oscillator and the radiation exciting it is in fact always an irreversible process, which essentially consists 
in compensating over time all initially existing spatial and temporal fluctuations in the radiation intensity.
When the stationary state has finally been reached, the energy of an oscillator of natural frequency $\nu$ and any 
damping decrement has the value$\,$\footnote{$\:${\color{red}{\it(M. Planck)}} op. cit. (Sitzgsber. Berl. Akad. Wiss. of 18 V. 1899) equation (34).}:
\vspace*{-4mm}
\begin{align}
U \; = \; \frac{c^2}{\nu^2} \:.\; R_{\,\nu} \: ,
\label{Eq_1}
\end{align}
where $c$ is the speed of light and 
$R_{\,\nu} \:.\: dv \:.\: d\sigma \:.\: d\Omega \:.\: dt$  
is the amount of energy that a linearly polarised beam transmits within the spectral region $d \nu$, through any surface element $d\sigma$ located in the transmitted vacuum perpendicular to it, within the opening angle $d\Omega$, and in the time $dt$. 
The essence of this equation, which has provided me with indispensable services, is that according to it the energy of the resonating oscillator depends only on the radiation intensity $R_{\,\nu}$ and its oscillation frequency $\nu$, but not on its other properties.
As a result of the irreversibility of these processes, it is now easy to specify a state function whose value always increases with time and which can therefore be interpreted as entropy. 
The entropy of the entire structure under consideration is made up of the sum of the entropies of all oscillators and the entropy of the cavity radiation. 
For the entropy of an oscillator, I set$\,$\footnote{$\:${\color{red}{\it(M. Planck)}} op. cit. (Sitzgsber. Berl. Akad. Wiss. of 18 V. 1899) equation (41).} 
\begin{align}
S \; = \; - \: \frac{U}{a\:\nu} \;.\; 
  \ln\left( \frac{U}{e\:b\:\nu} \right) 
   \; ; \;\;\;
   {\color{red}\mbox{\it or equivalently (P. Marquet):}\;
   \left( \frac{U}{a\:\nu} \right) \,.\,
   \left[\:
   1 \: - \: 
   \ln\left( \frac{U}{b\:\nu} \right) 
   \:\right] } 
\label{Eq_2}
\end{align}
where $a$ and $b$ are two universal constants and $e$, the basis of the natural logarithms, is only added as a factor to the constant $b$ for convenience
{\color{red}{\it(see the second formulation of (\ref{Eq_2}) without the term ``$e$'' / P. Marquet)}}, 
while the expression of the entropy of the cavity radiation results analogously from the assumption that 
every beam carries with it
a corresponding entropy
along 
with its energy, whereby a spatial entropy density can then be determined analogously to the spatial energy density.

On the basis of these determinations, I was able to prove that the entropy of the entire structure increases with time for any chosen initial state of both the oscillators and the cavity radiation. 
The final stationary state, that of thermodynamic equilibrium, in which the entropy reaches its maximum, depends in all its parts on a single parameter $T$, which is given by the relationship 
\begin{align}
\frac{dS}{dU} \; = \; \frac{1}{T} \: ,
\label{Eq_3}
\end{align}
and which therefore, thermodynamically speaking, denotes the absolute temperature.
If the value of $S$ from (\ref{Eq_2}) is substituted in this equation and the relationship (\ref{Eq_1}) is taken into account, the result for the radiation intensity of the oscillation frequency $\nu$ is
\vspace*{-1mm}
\begin{align}
R_{\,\nu} \; = \; \frac{b\:\nu^3}{c^2} 
\;.\; \exp\left( -\:\frac{a\:\nu}{T} \right)
\: .
\label{Eq_4}
\end{align}
This is the law of normal energy distribution established by W. Wien back in 1896, which was essentially confirmed by all the measurements available at the time (May 1899). 
So everything seemed to be in satisfactory order.

But soon afterwards, first O. Lummer and E. Pringsheim, and later F. Paschen, drew attention to certain deviations from Wien's distribution law, which they had found when extending their experiments to longer wavelengths and which became so clear in the course of the steadily increasing accuracy of the measurements that serious doubts had to arise about the general validity of the formula (\ref{Eq_4}). This prompted me to check whether the expression (\ref{Eq_2}) of the entropy of an oscillator could be replaced by a better one.
In my in-depth study of this problem, 
destiny made 
that an external circumstance that I had previously found unpleasant --the lack of interest among my fellow experts in the research direction I was pursuing-- now came as a certain relief to my work. At that time, a number of outstanding physicists had turned their attention to the problem of energy distribution in the normal spectrum, both experimentally and theoretically. But all of them were only looking in the direction of representing the radiation intensity $R_{\,\nu}$ as a function of the temperature $T$, while I suspected the deeper connection in the dependence of the entropy $S$ on the energy $U$. Since the importance of the concept of entropy had not yet been recognised at that time, nobody cared about the method I used, and I was able to carry out my calculations with leisure and thoroughness without having to fear any interference or revision 
from any side.

In order to gain a deeper insight into the properties of entropy, I first calculated in general terms, without making use of the relationship (\ref{Eq_2}), the entropy change that occurs overall when an oscillator located in a stationary radiation field, whose energy exceeds its value corresponding to the radiation field by a small amount $\Delta U$, absorbs the energy $d U$ from the radiation field.
This entropy change resulted 
in$\,$\footnote{$\:${\color{red}{\it(M. Planck)}} Ann. Physics I, 730 (1900).}:
\vspace*{-1mm}
$$ \frac{3}{5} \:.\: \frac{d^2S}{dU^2} 
   \:.\: \Delta U \:.\: d U \; . $$
Since $dU$ and $\delta U$ have opposite signs when a change actually occurs in nature, and since the above expression is always positive according to the second heat theorem, the following necessarily follows: 
$$ \frac{d^2S}{dU^2} \; < \; 0 \; . $$
In fact, the expression (\ref{Eq_2}) of entropy, which leads to Wien's distribution law, yields:
\begin{align}
\frac{d^2S}{dU^2} 
\; = \; 
-\:\frac{1}{a\:\nu\:U} 
\: .
\label{Eq_5}
\end{align}

The striking simplicity of this relationship suggested to me the idea of deriving it directly by means of a suitable descriptive consideration. I did so, and in this way arrived at the relationship (\ref{Eq_2}) from a different angle, and thus at Vien's distribution law. However, I refrain from reproducing it here because the reasoning is reasonably plausible, but by no means captivating. 
The fact that it is not true 
in reality results from the fact that Wien's distribution law is not generally confirmed by the measurements$\,$\footnote{$\:$O. Lummer. und E. Pringsheim, Sitzgsber. dtsch. physik. Ges. {\bf 2}, 163 (1900).}.  
So my attempts to improve formula (\ref{Eq_2}) had reached a 
dead point, 
and I was about to abandon it for good.
Then an event occurred which was to bring about a decisive change in this matter. At the meeting of the German Physical Society on 19 October 1900, F. Kurlbaum presented the results of the energy measurements he had carried out in collaboration with H. Rubens for very long wavelengths, which showed, among other things, that with increasing temperature the radiation intensity of the black body becomes more and more approximately proportional to the temperature $T$, in flagrant 
contrast to Wien's distribution law (\ref{Eq_4}), according to which the radiation intensity should always remain finite. 
Since this result had already become known to me a few days before the meeting through verbal communication from the authors, I had time to draw the conclusions from it in my own way before the meeting, and to utilise it to calculate the entropy of a resonating oscillator.
If for high temperatures $T$ the radiation intensity $R_{\,\nu}$ is proportional to the temperature, then according to (\ref{Eq_1}) the energy of the oscillator is also proportional to it, 
i.e.:$\,$\footnote{\label{footnote_U0_a}$\:${\color{red}{\it I have introduced the integration constant $U_0$ that is implicitly set to $C\:T_0$ by Max Planck to cancel the auxiliary constant terms and to arrive at $\:U = C \: T$ (P. Marquet)}}.}  
$$  U \;=\; C \:.\: T \; ,
    {\color{red} \;\;\;\;\;\;\; 
     \mbox{in fact:} \;\;\;
    U \;=\;C \:.\: T \;\;
    \;+\; \left[\: 
      U_0 \;-\; C \:.\: T_0  \;=\; 0 \, ?
    \:\right] 
    \;\;\;\;\;\mbox{(P. Marquet)}
    }
$$ 
and from this according to (\ref{Eq_3}) and by integration
{\color{red}\it(of $dS=dU/T$ with $dU=C\:dT$)}:$\,$\footnote{\label{footnote_S0_a}$\:${\color{red}{\it Note that the differential relationship $dS=dU/T$ with $dU=C\:.\:dT$ indeed leads to $dS = C \:.\: dT/T = C \:.\: d\ln(T)$ provided that $C$ is a constant, and thus to $S = C \:.\: \ln[\:(C\:T)/(C\:T_0)\:] + S_0$  after integration, with therefore the integration constant $S_0$ implicitly set to $C\:\ln(C\:T_0)$ by Max Planck, in order to cancel this auxiliary constant terms and to arrive at the relationship $S=C\:ln(U)$ with the use of the  hypothesis $U=C\:.\:T$ (P. Marquet)}}.} 
$$
 S \;=\; C \:.\: \ln(U) \; .
    {\color{red} \;\;\;\;\;\;\; 
    \mbox{in fact:} \;\;\;
    S \;=\; C \:.\: \ln(U)
    \;+\; \left[\: 
      S_0 \;-\; C \:.\: \ln(\,C\:T_0\,) \;=\; 0 \, ?
    \:\right]
    \;\;\;\;\;\mbox{(P. Marquet)}
    }
$$
Consequently:
\vspace*{-4mm}
\begin{align}
\frac{d^2S}{dU^2} 
\; = \; 
-\:\frac{C}{U^2} 
{\color{red} \;\;\;\;\;\;\;\;\; 
\mbox{or:} \;\;\;
\frac{d^2S}{dU^2} 
\; = \; 
-\:\frac{1}{U^2/C}
\;\;\;\;\;\mbox{(P. Marquet)}
} 
\: .
\label{Eq_6}
\end{align}
For large values of $U$, this relationship therefore replaces the relationship (\ref{Eq_5}), which is valid for small values of $U$. If we are now looking for a more general relationship that contains the two aforementioned (\ref{Eq_5}) and (\ref{Eq_6}) as limiting cases, the following is the simplest: 
$$
\frac{d^2S}{dU^2} 
\; = \; 
-\:\frac{1}{a\:\nu\:U \:+\: U^2/C} 
{\color{red} \;
\; = \; 
\frac{1}{a\:\nu} 
\;.\: 
\left[\:
  \frac{1}{U+a'\:\nu} \:-\: \frac{1}{U} 
\:\right]}
\; ,
$$ 
and by 
integration$\,$\footnote{\label{footnote_T0_a}$\:${\color{red}{\it Note that I have added (in red) additional intermediate formulations in (\ref{Eq_7}) and in the preceding equation, in order to better indicate how the integration must be achieved.
Note also that an additive integration constant $1/T_0(\nu)$ (in red and a priori depending on $\nu$ via the term $U_0+a'\:\nu$) appears in the r-h-s of (\ref{Eq_7}), with therefore the assumption $1/T_0(\nu)=0$ implicitly made by Max Planck in (\ref{Eq_7}) for all frequencies (P. Marquet)}}.} 
\begin{align}
\!\!\!\!
\frac{dS}{dU} 
\; = \; 
\frac{1}{T} 
{\color{red} \;
\; = \; 
\frac{1}{a\:\nu} 
\;.\: 
\left[\:
\ln\left( U + a'\:\nu \right)
\:-\:
\ln\left( U \right)
\:\right]}
{\color{red} 
\;+\;
\frac{1}{T_0(\nu)}}
\; = \; 
\frac{1}{a\:\nu} 
\;.\: 
\ln\left(
1 \:+\:\frac{a'\:\nu}{U}
\right)
{\color{red} 
\:+\;
\left(\frac{1}{T_0(\nu)} \equiv 0 \, ? \right)}
\: ,
\label{Eq_7}
\end{align}
where the constant $a \: C = a'$ is set as an abbreviation.

This is, if one reintroduces $R_{\,\nu}$ for $U$ according to (\ref{Eq_1}), the formula for the energy distribution law which, converted to wavelengths, I presented and recommended for examination in the course of the lively discussion following the Kurlbaum's lecture at the above-mentioned meeting of the German Physical Society.$\,$\footnote{$\:${\color{red}{\it(M. Planck)}} Sitzgsber, dtsch, physik. Ges. {\bf 2}, 202 (1900).}

On the morning of the next day, my colleague Rubens came to see me and told me that, after the conclusion of the meeting that very night, he had compared my formula exactly with his measurement data and found a satisfactory agreement everywhere. Lummer and Pringsheim, who initially thought they had found 
deviations$\,$\footnote{$\:$M. v. Laue, Natuwiss. 29, 137 (1941).}, soon withdrew their objection because, as Pringsheim told me verbally, it turned out that the deviations they had found were caused by a calculation error. The formula (\ref{Eq_7}) was then repeatedly confirmed by later measurements, the more precisely the more refined the experimental methods worked.$\,$\footnote{$\:$H. Rubens und G. Michel, Physik. Z. 22, 569 (1921).}
\vspace*{0mm}


\section{\underline{Section-III} {\color{red}{\it(The theoretical formula for Radiation)}} (p.157-159)} 
\vspace*{-2mm}

Thus the question of the law of spectral energy distribution in the radiation of the black body could be regarded as finally settled.
But now the theoretically most important problem remained: to give a proper justification of this law, and this was an incomparably more difficult task, because it required a theoretical derivation of the expression of the entropy of an oscillator, as it results  from (\ref{Eq_7}) by integration.
It can be written in the following 
form:$\,$\footnote{\label{footnote_T0_S0a}$\:${\color{red}{\it Note that the integration of (\ref{Eq_7}) is made between $U_0$ and $U$, with the frequency $\nu$ considered as a constant.
Accordingly I have included (in red) in the second line of (\ref{Eq_8}) the impacts of both the previous additive integration constant $1/T_0(\nu)$ in the r-h-s of (\ref{Eq_7}) and of a new global integration constant $S'_0(\nu)$ depending on the frequency via the terms $U_0+a'\:\nu$. 
It is thus needed to set $1/T_0(\nu)=0$ and $S'_0(\nu)=0$ for all frequencies in order to cancel the second line of (\ref{Eq_8}) and to arrive at the Planck's formula made of the first line only. 
(P. Marquet)}}.}
\vspace*{0mm}
\begin{align}
S{\color{red}\:(U, \, \nu)} & \: = \; 
\left(\frac{a'}{a}\right)
\left[\:
     \left(\frac{U}{a'\:\nu}\:+\:1\right) \:
\ln\!\left(\frac{U}{a'\:\nu}\:+\:1\right)
\;-\:\left(\frac{U}{a'\:\nu}\right) \:
\ln\!\left(\frac{U}{a'\:\nu}\right)
\:\right] 
\; .
\label{Eq_8} \\
 & \quad \quad
  {\color{red} 
   + \; 
 \left\{\:
    S'_0(\nu)
     \, + \,  
   \frac{\left(\,U\,-\,U_0\,\right)}{T_0(\nu)}
   \; \equiv \; 0 \, ?
  \:\right\} }
\nonumber \: 
\end{align}
\dashuline{\,In order to give this expression a physical meaning}, completely \dashuline{\,new considerations} about the nature of entropy \dashuline{were necessary}, which go beyond the field of electrodynamics.


Among the physicists of the time, Ludwig Boltzmann was the one who understood the meaning of entropy most profoundly. He interpreted the entropy of a physical entity in a certain state as a measure of the probability of this state, and he saw the content of the second law in the fact that, for every change that occurs in nature, the  structure changes into a more probable state.
In fact, he had succeeded in defining a state function $H$ {\it\color{red}(Eq.~8, p.33)} in his kinetic theory of gases,$\,$\footnote{$\:$L. Boltzmann, Vorlesungen über Gastheorie, I. Teil. Leipzig: Johann Ambrosius Barth 1896, S. 33.} which has the property of decreasing in magnitude with every change of state that occurs in nature, and which can therefore be regarded as the negative entropy. However, in order to prove this famous $H$ theorem, he had to resort to the restrictive hypothesis that the state of the gas is ``\,molecularly disordered.\,''

Until then, 
I had not paid any attention myself to
the connection between entropy and probability.
This 
was not tempting 
for me because every law of probability also allows for exceptions, and because at the time I ascribed validity to the second heat theorem without exception. 
That the proof of the irreversibility of the radiation processes I considered could only succeed on the condition of the hypothesis of ``natural radiation'' (i.e. that such a restrictive hypothesis is just as necessary in the theory of radiation and plays the same role there as that of molecular disorder in the theory of gases)  only became completely clear to me with time.

But since there was no other way out for me, I tried Botzmann's method and generally 
set$\,$\footnote{{\color{red}{\it$\:$Note that, like in the footnote \ref{footnote_S0_a}, Planck did not write the arbitrary additive reference values (or integration constant) $S''_0$ (in red) in (\ref{Eq_9}), or said differently Planck assumed that $S''_0=0$ leading to the expected result that: $S=0$ and $W=1$ for the dead-state defined at $T=0$~K and for which no more degree of freedom exist (and thus with only $W_0=1$ possible state at $T=0$~K): this corresponds to the Third Law of thermodynamics, gradually defined by Max Planck in the editions from 1911 to 1917 of his textbooks about Thermodynamic and Radiation (P. Marquet)}}.} 
\begin{align}
S \; = \; k \:.\; \ln(W) 
  {\color{red} \:\;+\;\, (\:S''_0 \;\equiv\; 0?\:)} \:\: ,
\label{Eq_9}
\end{align}
for any state of any physical structure, where $W$ denotes the appropriately calculated probability of the state designated.
If this relationship is really to have general significance, then, since entropy is an additive quantity but probability is a multiplicative quantity, \dashuline{\,the constant $k$ must be a universal number\,} that depends only on the units of measurement. Understandably, it is often referred to as Boltzmann's constant. However, it should be noted that \dashuline{\,Boltzmann neither ever introduced this constant nor\,}, to my knowledge, ever \dashuline{\,thought of asking for its numerical value\,}. For then he would have had to deal with the number of real atoms --a task that he left entirely to his colleague J. Loschmidt, while he himself always kept in mind in his calculations the possibility that the kinetic theory of gases only represents a mechanical picture. It was therefore sufficient for him to stop at the larger atoms.

In order to apply the relationship (\ref{Eq_9}) to the present case, I thought of a structure consisting of a very large number $N$ of completely similar oscillators and tried to calculate the probability that this structure has the given energy $U_N$. Since a probability quantity can only be found by counting, it was above all necessary to regard the energy $U_N$ as a sum of discrete, identical elements $\varepsilon$, the number of which may also be denoted by the very large number $P$.

Therefore: 
\vspace*{-4mm}
\begin{align}
U_N \; = \; N \,.\; U \; = \; P \,.\; \varepsilon \: ,
\label{Eq_10}
\end{align}
where $U$ means the average energy of an oscillator.

Then the number of different ways in which the $P$ energy elements can be distributed among the (imagined numbered) $N$ oscillators$\,$\footnote{$\:${\color{red}{\it(M. Planck)}} Sitzgsber. dtsch, physik. Ges. vom 14. XII. 1900, S. 240.} is:
\vspace*{-3mm}
\begin{align}
W \; = \; \frac{(\,P\,+\,N\,)\,!}{P\,! \;\; N\,!} \: .
\label{Eq_11}
\end{align}
From this, according to (\ref{Eq_9}), the entropy of the oscillator system is: 
$$ S_n \; = \; N \,.\; S  \; = \; 
k \:.\; \ln\left[\: 
\frac{(\,P\,+\,N\,)\,!}{P\,! \;\; N\,!} 
\:\right] \: ,$$
and according to Stirling's theorem$\,$\footnote{$\:${\color{red}{\it Namely: $\ln(n\,!) \approx n\,\ln(n) - n$ (P. Marquet)}}.}: 
$$ N \,.\; S \; = \; 
k \:.\: \left\{\:
(P+N)\:\ln(P+N) \:-\: P\:\ln(P) \:-\: N\:\ln(N)
\:\right\} \: ,$$ 
or: 
\vspace*{-4mm}
\begin{align}
S \; = \; 
k \:.\: \left\{\:
\left(\frac{P}{N}+1\right)\:\ln\!\left(\frac{P}{N}+1\right) 
\:-\: 
\left(\frac{P}{N}\right)\:\ln\!\left(\frac{P}{N}\right) 
\:\right\}
{\color{red}\:\;+\;\, (\:S''_0 \;\equiv\; 0?\:)} 
\:\: .
\label{Eq_12}
\end{align}
The similarity between the two expressions (\ref{Eq_8}) and (\ref{Eq_12}) is 
obvious.$\,$\footnote{$\:${\color{red}{\it To do so, it is explained in the footnote~\ref{footnote_T0_S0a} that the second  line of (\ref{Eq_8}) must be discarded and set to zero through a suitable set of the two constants $S'_0(\nu)=0$ and $1/T_0(\nu)=0$ for all frequencies.
Moreover it is needed to set $S''_0=0$ in (\ref{Eq_12}) in order to allow a clear identification of (\ref{Eq_8}) with (\ref{Eq_12}) and to allow the writing of the next relationship $P/N=U/(a'\:\nu)$, which depends on the frequency $\nu$
(P. Marquet)}}.}

The only thing left to do is to make the necessary determinations to make them completely identical. This is done by setting: 
\vspace*{-2mm}
$$ \boxed{\:k \; = \; \frac{a'}{a}\:} 
   \;\;\;\;\;\; \mbox{and} \;\;\;\;\;\; 
   \boxed{\:\frac{P}{N}\; = \; \frac{U}{a'\:\nu}\:} \; . $$
According to (\ref{Eq_10}), the quantity of the energy element follows from this: $\varepsilon = a'\:\nu$. I denoted the constant $a'$, which is independent of the nature of the oscillators, by $h$ and, since it has the dimension of a product of energy and time, I called it the \dashuline{\,elementary quantum of action\,} or the \dashuline{\,action element\,}, in contrast to the \dashuline{\,energy element\,} $h\:\nu$. With the measured values of the constants $a$ and $a'$ of the radiation law (\ref{Eq_7}), the following values of $k$ and $h$ were obtained:
$$ k \; = \; 1.346 \:.\:10^{-16} \: \mbox{erg\,/\,grad}
   \;,\;\;\;\;
   h \; = \; 6.55 \:.\:10^{-27} \: \mbox{erg\,.\,sec} 
   \;\; . \;\;\;\;\;
   \mbox{{\color{red}{\it(with $1$~erg$\:=10^{-7}$~J 
                         / P. Marquet)}}}
$$

As far as the experimental testing of this theory was concerned, this was initially only possible to a very limited extent, because only the first constant $k$ was available for this purpose, the numerical value of which was at most reasonably known in terms of magnitude. 
Moreover, differently to Boltzmann$\,$\footnote{$\:$L. Boltzmann, Sitzgsber. Wien. Akad. Wiss. (II) {\bf 76}, 428 (1877).},  
this means that the so-called absolute gas constant $R = 8.31\:.\:10^7\: \mbox{erg/grad}$ is no longer in relation to large molecules, but in relation to the actual molecules.

Therefore, the ratio $k/R = 1.62\:.\:10^{-24}$ {\color{red}{\it(namely $1/6.17\:.\:10^{24}$, the reverse of the Avogadro's number)}} is the reduction factor, which reduces the mass of a large molecule to the mass of the real molecule. From this I also calculated the value of the electric elementary quantum by multiplying the reduction factor by the charge $2.895\:.\:10^{14}$ (electrostatic) of a monovalent large ion, \dashuline{\,to get $4.69\:.\:10^{-10}$\,} (electrostatic), while F. Richarz calculated $1.29\:.\:10^{-10}$ and J. J. Thomson had found $6.5\:.\: 10^{-10}$. No further measurements of the electric elementary quantum were available at that time.

So I could be reasonably satisfied with this result. In the physical world, however, things looked somewhat different. The calculation of the electric elementary quantum from heat radiation measurements was not even taken seriously in some places. 
But I did not allow myself to be misled by such doubts in my confidence in my constant $k$. 
However, I only gained complete certainty when I learnt that E. Rutherford and H. Geiger had \dashuline{\,arrived at the value $4.65\:.\:10^{-10}$\,} by counting $\alpha$ particles. Since then, refined measurement methods have \dashuline{\,led to a small increase in this number\,} {\color{red}{\it(and thus with a good agreement with the Planck's value of $\,4.69\:.\:10^{-10}$/ P. Marquet)}}.

The task of experimentally testing the numerical value of the second constant, $h$, which was initially completely up in the air, seemed much more hopeless. 
It was therefore a great surprise and pleasure for me when Ms J. Franck and G. Hertz, in their experiments on the excitation of a spectral line by electron collisions, found a method of measurement that could not have been more direct. This also removed the last doubt about the reality of the quantum of action.
Now, however, the theoretically most difficult problem arose of assigning a physical meaning to this strange constant {\color{red}{\it(i.e. $h$)}}, because its introduction meant a break with classical theory that was much more radical than I had initially assumed. The nature of entropy as a measure of probability in the sense of Boltzmann had also been definitively established for radiation. This became particularly clear in a theorem whose validity the closest of my students, Max v. Laue, convinced me of in several conversations: that the entropy of two coherent radiation beams is smaller than the sum of the entropies of the individual beams, entirely in accordance with the theorem that the probability of the simultaneous occurrence of two interdependent events is different from the product of the probabilities of the individual events.
But the nature of the energy elements $h\:\nu$ remained unclear. \dashuline{\,For several years, I repeatedly tried to somehow incorporate the quantum of action into the system of classical physics. But I did not succeed\,}. Rather, the development of quantum physics was, as is well known, reserved for younger forces, of whom I only mention here the names of A. Einstein, N. Bohr, M. Born, P. Jordan, W. Heisenberg, L. de Broglie, E. Schrödinger, P. A. M. Dirac, while the mathematical structure of the theory among German physicists was primarily the work of A. Sommerfeld and the promotion of physical understanding made by CL. Schaefer.

At  the current stage of development, it is probably not yet possible to say with certainty whether we have already reached a final stopping point, as many outstanding physicists assume. I am not one of them. Rather, I believe that fundamental changes in our physical conceptualisation, which cannot yet be clearly foreseen, are still required before quantum theory reaches the same degree of perfection as classical theory did in its day.
\vspace*{2mm}

\end{document}